\documentclass[aps, pra, superscriptaddress, reprint, twocolumn, a4, footinbib]{revtex4-1}
\usepackage{graphicx}
\usepackage{amsmath,amssymb}
\usepackage{dcolumn}
\usepackage{bm}
\usepackage{physics}
\usepackage{dsfont}
\usepackage{mathrsfs}
\usepackage{bbold}

\newcommand{\comment}[1]{}

\newcommand{\ns}{\mathrm{\;ns^{-1}}}

\begin{document}
	\title{Protocol for generating multi-photon entangled states from quantum dots in the presence of nuclear spin fluctuations}
\date{\today}

\author{Emil V. Denning}
\affiliation{Department of Photonics Engineering, DTU Fotonik, Technical University of Denmark, Building 343, 2800 Kongens Lyngby, Denmark}

\author{Jake Iles-Smith}
\affiliation{Department of Photonics Engineering, DTU Fotonik, Technical University of Denmark, Building 343, 2800 Kongens Lyngby, Denmark}

\author{Dara P. S. McCutcheon}
\affiliation{Quantum Engineering Technology Labs, H. H. Wills Physics Laboratory and Department
of Electrical and Electronic Engineering, University of Bristol, BS8 1FD, UK}

\author{Jesper Mork}
\email{jesm@fotonik.dtu.dk}
\affiliation{Department of Photonics Engineering, DTU Fotonik, Technical University of Denmark, Building 343, 2800 Kongens Lyngby, Denmark}

\begin{abstract}
Multi-photon entangled states are a crucial resource for many applications in quantum information science. 
Semiconductor quantum dots offer a promising route 
to generate such states by mediating photon-photon correlations via a confined electron spin, but 
dephasing caused by the host nuclear spin environment  
typically limits coherence (and hence entanglement) between photons to the spin $T_2^*$ time of a few nanoseconds. 
We propose a protocol for the deterministic generation of 
multi-photon entangled states that is inherently robust against 
the dominating slow 
nuclear spin environment fluctuations, meaning that coherence and entanglement 
is instead limited only by the much longer spin $T_2$ time of microseconds. 
Unlike previous protocols, the present scheme 
allows for the generation of very low error probability polarisation encoded 
three-photon GHZ states and larger entangled states, without the need for 
spin echo or nuclear spin calming techniques. 
\comment{Do we think we can call out scheme `deterministic' given the frequency eraser issue?}
\comment{I'm not sure what we mean by the external field retaining the QD 
eigenstructure in the presence of nuclear spin fluctuations.} 

\end{abstract}
\maketitle

\section{Introduction}
A crucial requirement for photonic measurement-based quantum computing schemes is a resource of entangled 
states~\cite{ladd2010quantum,nielsen2010quantum,gisin2002quantum,ekert1991quantum,bennett1993teleporting,wang2012quantum,raussendorf2001one,raussendorf2003measurement,briegel2009measurement,kok2007linear}. 
The generation of such states is being pursued on various platforms; among 
these are continuous variable quantum optics~\cite{yokoyama2013ultra}, spontaneous parametric 
down-conversion in nonlinear crystals~\cite{kwiat1999ultrabright},   
nitrogen-vacancy centres~\cite{togan2010quantum}, and self-assembled semiconductor quantum 
dots (QDs)~\cite{schwartz2016deterministic}. 
QDs in particular are attractive due to the combination of their 
excellent optical properties~\cite{claudon2010highly,gazzano2013bright, somaschi2016near, he2013demand,Iles-Smith2017Nature}, 
and the prospect of deterministic interactions with single photons~\cite{arcari2014near,claudon2010highly}. 
By charging a QD with a single electron, it becomes equipped with an internal 
spin degree of freedom that couples to the polarisation of optical photons~\cite{bayer2002fine}, 
while also benefiting from highly developed optical control and readout techniques~\cite{atature2006quantum,xu2007fast,berezovsky2008picosecond,press2008complete,kim2010fast,delteil2014observation,carter2013quantum,sun2016single,gao2015coherent}. 
Using these properties, it is possible to 
generate spin--photon entanglement~\cite{hu2008deterministic,PhysRevB.88.035315}, 
and by entangling a sequence of photons with a QD,  
spin--multi-photon states are generated, reducing to multi-photon entangled states once the QD spin is measured~\cite{lindner2009proposal,economou2010optically,buterakos2017deterministic}.

A considerable challenge for the QD platform is posed by the interaction of the QD spin with its nuclear spin environment, 
which gives rise to a slowly fluctuating magnetic Overhauser field~\cite{mccutcheon2014error,merkulov2002electron}. 
Due to uncertainty in the 
Overhauser field, 
phase coherence between the QD spin states is lost on a timescale 
set by the spread of available Overhauser states, limiting 
the QD spin coherence to 
typically only a few nanoseconds~\cite{press2010ultrafast,urbaszek2013nuclear,kuhlmann2013charge} 
(usually termed the $T_2^*$, ensemble, or inhomogeneous dephasing time). 
This renders practical implementations to generate states beyond spin--single photon entanglement 
extremely challenging in their original formulations~\cite{lindner2009proposal,hu2008deterministic,PhysRevB.88.035315,economou2010optically,buterakos2017deterministic,leuenberger2014deterministic}. 
Spin coherence times can in principle be extended beyond $T_2^*$ 
by applying spin echo or dynamical decoupling sequences which unwind fluctuating phase evolution~\cite{press2010ultrafast,vu2012real}.  
However, this not only adds operational complexity, but in cases which utilise 
photon frequency degrees of freedom~\cite{gao2012observation}, will not extend photon coherence times, 
as the Overhauser field is imprinted onto the photonic component of the state not affected by echo pulses.
Spin coherence may also be extended by polarisation of the nuclear environment~\cite{ethier2017improving,urbaszek2013nuclear,chekhovich2013nuclear, maletinsky2007dynamics,bracker2005optical,imamoglu2003optical,lai2006knight}, 
though a very high ($>90\%$) and as yet unachievable degree of polarisation is required.

\section{Dephasing-resilient protocol}
As a solution to this, we propose a QD-based protocol to generate multi-photon entangled states 
that is naturally robust against slow Overhauser field fluctuations, 
with the coherence being instead limited only by faster pure-dephasing 
(homogeneous) processes, with a typical timescale of microseconds (termed the $T_2$ time).
The central feature of our proposed protocol
is that it combines 1) an external field to ensure the nuclear environment 
gives rise to a fluctuating magnetic field amplitude only, 
with 2) narrow band excitation, which means an entangled state is generated 
in which all terms have the same energy. 
This means only a global inconsequential phase is acquired over time, 
thus ensuring robustness against the dominating slow nuclear spin fluctuations.
We benchmark our protocol against a multi-photon extension of the experimental 
realisations in Refs.~\cite{gao2012observation, schaibley2013demonstration, de2012quantum} 
and the theoretical schemes in Refs.~\cite{hu2008deterministic,lindner2009proposal}, 
showing that with realistic noise models these cannot 
be scaled to create entanglement beyond the spin--single photon regime 
as they lack one or both of the above properties.
Using the proposed protocol 
in combination with a suitable frequency quantum eraser, we show that three-photon 
GHZ states can be generated near deterministically with near-unity fidelity, and without 
any active measures taken to avoid nuclear spin dephasing.  
Several of these microclusters could then be efficiently transformed to a large cluster state using only passive linear optical elements \cite{gimeno2015three}.

\begin{figure}
	\centering
	\includegraphics[width=\columnwidth]
	{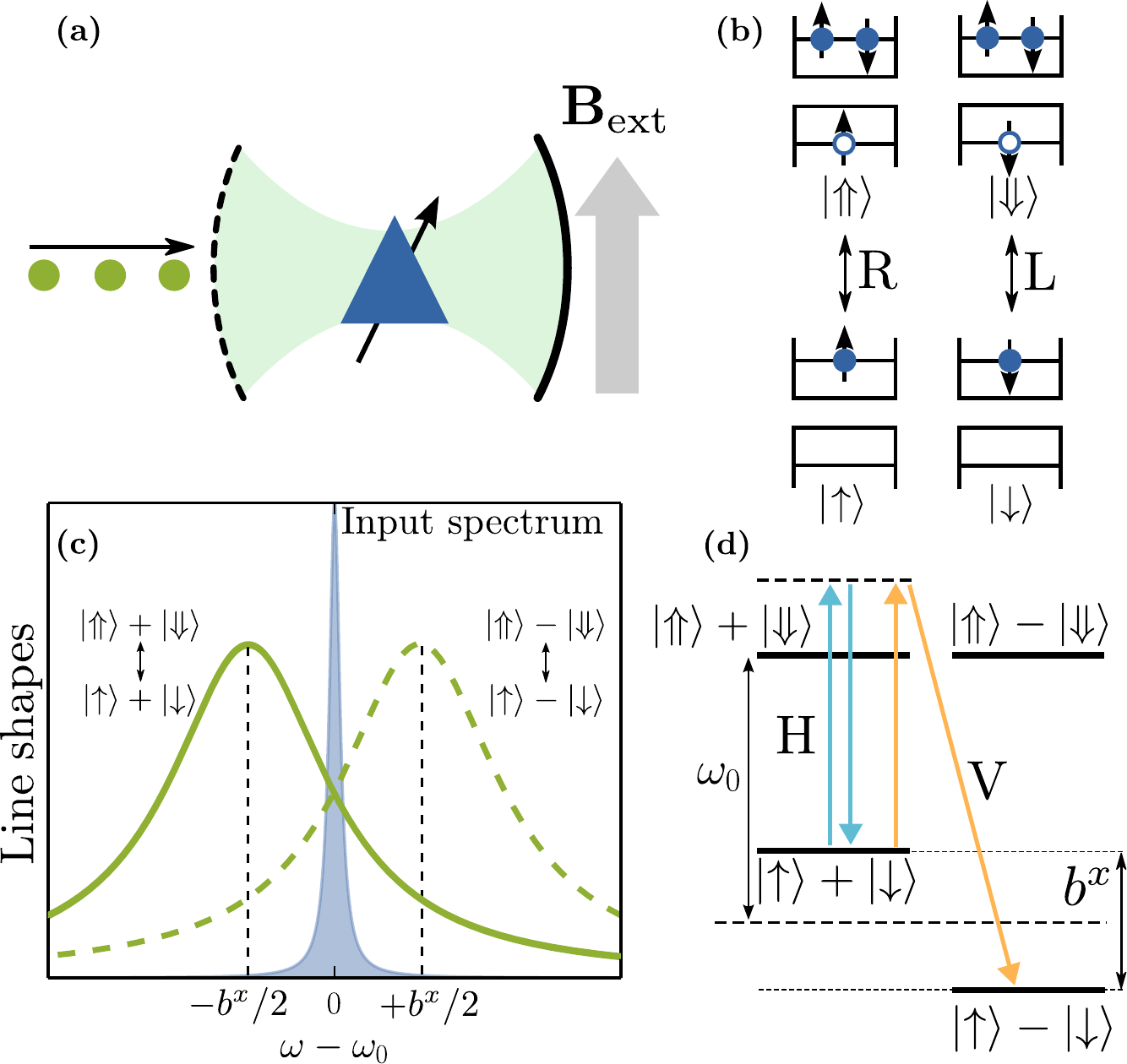}
	\caption{{\bf(a)} A QD in a
		polarisation-degenerate, single-sided cavity, is exposed to an
		external magnetic field perpendicular to the cavity axis. 
		{\bf (b)} Electron and hole configurations for the ground states and trions, 
		which in zero field are connected via circularly polarised transitions. 
		{\bf(c, d)} The Voigt-geometry magnetic field leads to linearly polarised transitions (labelled $H$ and $V$) 
		between hybridised levels as indicated, split by the Zeeman energy $b^x$. Shown in (c) is a spectrally narrow photon resonant with 
		the zero field transition energy $\omega_0$, which can lead to a spin-flip Raman scattering process changing the photon's energy and polarisation (orange arrows in (d)), 
		or a coherent scattering process leaving energy and polarisation unchanged (blue arrows). 
		Occurring in superposition these processes lead to spin-photon entanglement.}
	\label{fig:intro}
\end{figure}

Our protocol is based on a negatively charged QD in a single-sided, polarisation-degenerate cavity, 
operating in the weak coupling regime. An external magnetic field perpendicular to the optical 
axis splits the QD transitions, and results in linearly polarised transitions to the excited trion states. 
We now consider a $H$-polarised photon incident on the cavity, 
with the QD in the external magnetic field eigenstate $\ket{\phi_+}=(1/\sqrt{2})(\ket{\uparrow}+\ket{\downarrow})$, 
where $\ket{\uparrow}$ and $\ket{\downarrow}$ denote the ground state electron spin projection 
along the optical axis (defining the $z$-direction). 
If the incoming photon is resonant with the bare QD transition energy in zero field, labelled 
$\omega_0$, there are two off-resonant scattering possibilities.
A Raman transition can take place, in which the spin 
of the QD is flipped, and the photon frequency and polarisation are changed (orange arrows in Fig.~\ref{fig:intro}(d)), 
or the photon can coherently scatter, leaving 
it and the QD unchanged (blue arrows). As such, the composite QD--photon 
system will evolve in superposition, 
and we write a single photon scattering event as 
$\ket{H,\omega_0}_1\ket{\phi_+}\rightarrow \ket*{\psi^{(1)}}$ with 
\begin{align}
\label{eq:2qb-ket}
\ket*{\psi^{(1)}} \equiv  \frac{1}{\sqrt{2}}(\ket{H,\omega_0}_1\ket{\phi_+} 
-i\ket{V,\omega_+}_1\ket{\phi_-}),
\end{align} 
where $\omega_\pm = \omega_0 \pm (b^x/2)$ with $b^x$ is the Zeeman splitting, 
and $\ket{\alpha,\omega}_i$ denotes photon $i$ in polarisation state 
$\alpha$ with frequency $\omega$. The superscript on $\ket*{\psi^{(n)}}$ 
denotes the photon number in the scattered state.

A second photon can then be sent to the QD--cavity system after some time,  
and the total composite state will be the three-qubit entangled state (cf. App. \ref{sec:n-photon} for details)
\begin{align}
\label{eq:3qb-ket}
\ket*{\psi^{(2)}} = \!\tfrac{1}{2}\big( &\ket{H,\omega_0}_1\!\qty{\ket{H,\omega_0}_2\!\ket{\phi_+}   - i \ket{V,\omega_+}_2\!\ket{\phi_-}}
\\ + & \ket{V,\omega_+}_1\qty{-i\ket{H,\omega_0}_2\ket{\phi_-}
    +\ket{V,\omega_-}_2\ket{\phi_+}}\big).\nonumber
\end{align} 
This state is local unitary equivalent (LUE) to a three-qubit linear cluster state~\cite{raussendorf2003measurement} 
and a GHZ state, provided that the frequency degree of freedom is erased. For three or more photons, 
the state is no longer LUE to a GHZ or linear cluster state, though possesses a rich entanglement structure 
with maximal localisable entanglement and infinite entanglement length. 
Of particular note, when the QD spin is projected  
out of the state $\ket*{\psi^{(3)}}$ in the $\{\phi_{\pm}\}$ basis, 
the remaining state is LUE to a three-photon polarisation encoded GHZ state~\footnote{Alternatively, the polarisation information can 
be erased in order to achieve qutrit colour entanglement. However, we shall not discuss this possibility further in this paper.}.

The most important feature of Eq.~\eqref{eq:3qb-ket}, however, is that each term has the same total energy.
This is because the first Raman process flips 
the spin from $\ket{\phi_+}$ to $\ket{\phi_-}$, transferring energy $b^x$ from the QD 
to the photon. In the second spin-flip event, the opposite happens, and the 
photon transfers energy $b^x$ to the QD. 
Consequently, the state $\ket*{\psi^{(n)}}$ for any $n$ 
consists of 
a large superposition of trajectories that all share the same total energy $n\omega_0+b^x/2$, 
and $\ket*{\psi^{(n)}}$ will acquire only a global phase in time. 
Crucially, this means that when an ensemble of states such as $\ket*{\psi^{(n)}}$ is prepared, 
phase coherence between terms in the superposition is protected from any 
fluctuations in $b^x$ that may occur between one realisation and another. 
In particular, for a single QD, slow variations in the Overhauser field over time will 
not decohere $\ket*{\psi^{(n)}}$, allowing, for example, the generation of three-photon 
GHZ states with near-unit fidelity.

As this insensitivity to nuclear spin interactions 
is the essential feature of our protocol, 
we now consider it in more detail. 
The dominant coupling between the QD electron spin and 
nuclear spins is the
hyperfine interaction~\cite{cywinski2009electron}. 
If this is much weaker than the 
electron Zeeman energy and the number of nuclear spins is large, its effect 
can be modelled as a magnetic Overhauser field, $\mathbf{B}_\mathrm{N}$~\cite{merkulov2002electron}, 
which can be added to the external field to give
$\mathbf{B}=\mathbf{B}_\mathrm{ext}+\mathbf{B}_\mathrm{N}$. 
Due to the large number of nuclear spins 
$\mathbf{B}_\mathrm{N}$ evolves on a slow microsecond timescale, as compared to the characteristic 
nanosecond timescale governing the electron spin dynamics~\cite{merkulov2002electron}. 
This allows us to model the Overhauser field as being 
stationary during a single experimental run, 
but probabilistically chosen from  
$w(B_\mathrm{N}^i;\Delta_B)= 1/(\Delta_B\sqrt{2\pi})\exp[-(B_\mathrm{N}^i)^2/(2\Delta_B^2)]$, 
describing a Gaussian distribution with zero mean for each of the Cartesian components, 
$\smash{B^i_\mathrm{N}}$, and with standard deviation $\Delta_B$~\cite{merkulov2002electron}.
If the external field $\mathbf{B}_\mathrm{ext}=B_{\mathrm{ext}} \bf{\hat{x}}$ is appreciably stronger than $\Delta_B$, 
we can assume that the components of 
$\mathbf{B}_\mathrm{N}$ parallel to $\mathbf{B}_\mathrm{ext}$ dominate~\cite{merkulov2002electron}. 
In such a case nuclear spins can be included by writing the effective Zeeman splitting as 
$b^x=g_e\mu_B(B_\mathrm{ext}+B_\mathrm{N}^x)$, with 
$g_e$ the electron Land\'e factor and $\mu_B$ the Bohr magneton, 
and with $B_N^x$ averaged over using $w(B_\mathrm{N}^x;\Delta_B)$. 
\comment{Is it that the Zeeman energy needs to be larger than $A_i$ or $\mathcal{A}$?}

To see how ensemble dephasing can arise, consider 
a simple superposition state in the magnetic field eigenstate basis 
$(1/\sqrt{2})(\ket{\phi_+}+\ket{\phi_-})$. 
For times $t$ less than a microsecond, this state 
becomes $\ket{\varphi}=(1/\sqrt{2})(\mathrm{e}^{-i b^x t/2}\ket{\phi_+}+\mathrm{e}^{i b^x t/2}\ket{\phi_-})$ 
in a single realisation. An ensemble of such states, however, samples all Overhauser fields, 
giving the single-spin density operator 
$\varrho=\int\mathrm{d} B_N^x \, w(B_N^x;\Delta_B)\ket{\varphi}\bra{\varphi}$, 
and we find that coherences decay as 
$\bra{\phi_+}\varrho\ket{\phi_-}\propto\exp[-(t/T_2^*)^2]$ with 
$T_2^* =\sqrt{2} /(g_e \mu_B \Delta_B) $, 
which for typical InGaAs QDs corresponds to nanoseconds. 
Crucially, however, in our protocol, states such as $\ket{\varphi}$ above 
are never produced. Instead, 
assuming that all scattering processes take 
place within the microsecond time scale over which the Overhauser field can be considered constant, 
after accumulating $n$ photons in the 
composite state, it will have the form 
$\ket*{\psi^{(n)}}=(\ket*{\psi^{(n)}_+}\ket{\phi_+}+\ket*{\psi^{(n)}_-}\ket{\phi_-})/\sqrt{2}$, as we show in App.~\ref{sec:n-photon}. 
Here $\smash{\ket*{\psi_\pm^{(n)}}}$ is an entangled $n$-photon state, in which all terms have energy  
$\Omega_+=n\omega_0$ or $\Omega_-=n\omega_0+b^x$. 
This form 
eliminates the inhomogeneous ensemble dephasing as described above, as the phase 
can be factored out of the complete state. 
Dephasing only occurs on a much longer timescale of the $T_2$ time 
set by pure-dephasing processes, and typically corresponding to microseconds~\cite{merkulov2002electron}.  
If the spin is measured in the basis $\{\phi_+,\phi_- \}$ 
while the Overhauser field is unchanged, the photonic state is projected to one of the 
states $\smash{\ket*{\psi_\pm^{(n)}}}$, which are also robust against ensemble dephasing. 
Though we have emphasised resilience to Overhauser field fluctuations, 
by the same arguments our scheme is also robust against any other slow processes leading to 
energy level fluctuations, most notably those caused by charge noise~\cite{kuhlmann2013charge,reigue2017probing}.

Having shown that our protocol is robust against ensemble dephasing processes, 
we now turn our attention to another potential imperfection, that arising from 
photon scattering process itself, which we term the {\emph{scattering fidelity}}. 
We are interested here in a quantitative analysis of how well the entangled states 
in Eqs.~\eqref{eq:2qb-ket} and \eqref{eq:3qb-ket} are produced given a realistic QD--cavity model. 
To assess this, we write the total Hamiltonian as $\hat{H}=H_0(t) + H_B$, where $H_0(t)$ is the 
QD--cavity Hamiltonian including light--matter interactions, and $H_B$ contains 
the magnetic field. In a frame rotating at $\omega_0$ we have 
(we set $\hbar =1$) $H_0(t) = \eta(t)\mathbf{e}_\mathrm{in}^\dagger\mathbf{A} + g\mathbf{\Sigma}^\dagger\mathbf{A} + \mathrm{H.c}$, 
with $\mathbf{A}=(a_+, a_-)^\mathrm{T}$ the polarisation-resolved vectorial cavity mode operator 
in the circular polarisation basis, 
$\mathbf{\Sigma}=(\dyad{\uparrow}{\Uparrow}, \dyad{\downarrow}{\Downarrow})^\mathrm{T}$, 
and $g$ is the QD--cavity coupling strength.  The incoming light 
is modelled as a weak coherent pulse, described by a time-dependent driving of the cavity field, 
taken to be Gaussian, $\eta(t)=\eta_0\exp[-(t/t_0)^2]$, and $\mathbf{e}_\mathrm{in}$ is the input polarisation 
Jones vector in the circular basis. 
The magnetic field Hamiltonian is $H_B = \mu_B\mathbf{B}\cdot(g_e\mathbf{S}_e - g_h\mathbf{S}_h)$, 
with $\mathbf{S}_e$ ($\mathbf{S}_h$) the vectorial spin operator for the electron (hole) subspace and 
$g_h$ the hole Land\'e factor~\footnote{We restrict our analysis to the ideal case, where no hole mixing is present.}.
With a numerical solution of the dynamics generated by the Hamiltonian~\cite{johansson2013qutip}, 
the scattering fidelity for an $n$ photon state is simply 
$\mathcal{F}^{(n)}=\mathrm{Tr}[\rho \dyad*{\tilde{\psi}^{(n)}}]$, where 
$\rho$ is the numerically calculated QD-photon density operator, and 
$\dyad*{\tilde{\psi}^{(n)}}$ the ideal maximally entangled state~\footnote{We note that the frequency information is erased 
automatically in our computational model by projecting the photonic state onto a broad cavity quasimode. 
Thus, we calculate the fidelity in the limit of a perfect frequency erasure.}. Additional details about the dynamical model and calculation of fidelities can be found in Appendices~\ref{sec:model} and~\ref{sec:fidelity-measures}.

\begin{figure}
	\centering
	\includegraphics[width=\columnwidth]{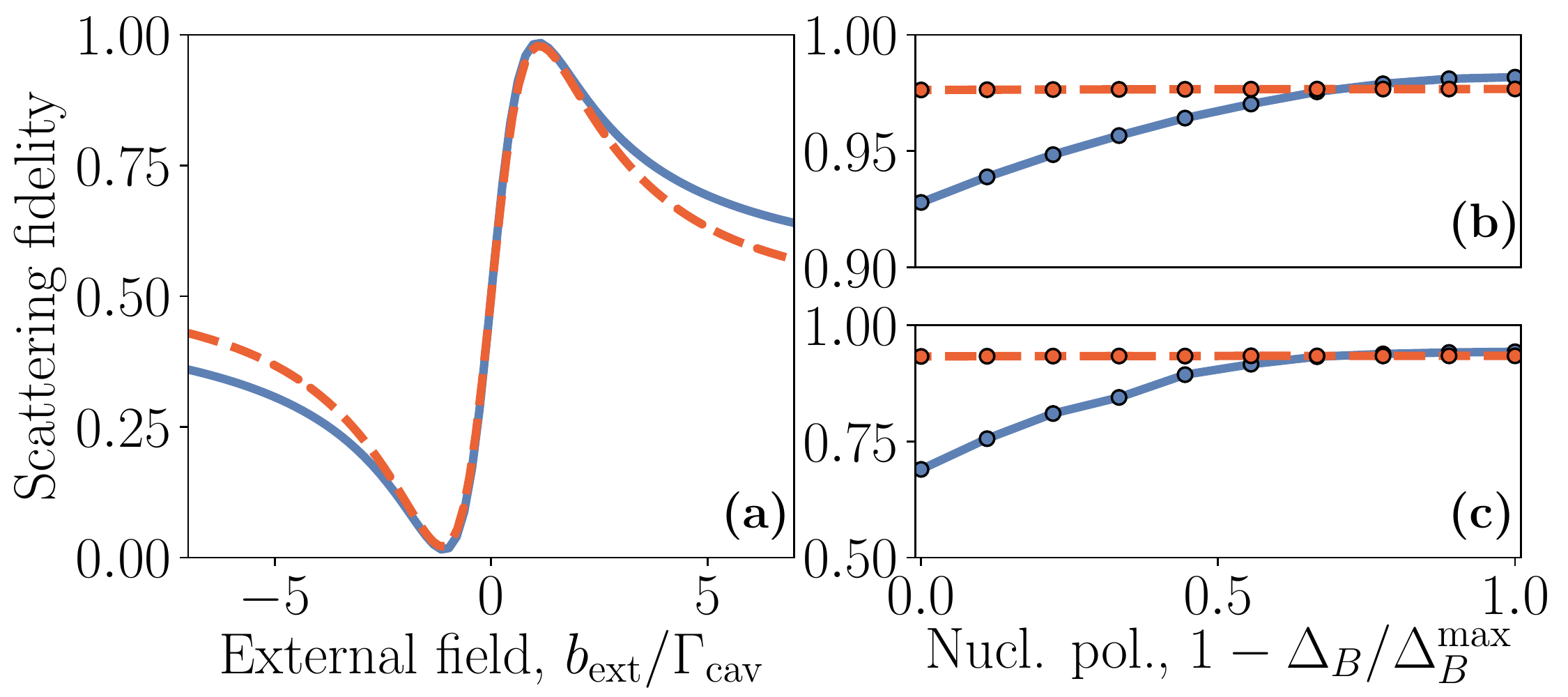}
	\caption{{\bf(a)} One photon scattering fidelity $\mathcal{F}^{(1)}$ with respect to the ideal Bell state, $(1/\sqrt{2})(\ket{x}\ket{\phi_+}-i\ket{y}\ket{\phi_-})$ 
	as function of external magnetic field in the absence of Overhauser field fluctuations. Blue solid lines correspond to a 
	low $Q$-factor cavity ($\kappa=10^3\ns,\; Q\simeq2000$); red dashed lines represent a high 
	$Q$-factor cavity ($\kappa=150\ns,\; Q\simeq13000$). Other parameters: $t_0 =
	8/\Gamma_\mathrm{cav}, \;g=15\ns,\; g_h/g_e = 0.2$, $\eta_0/\kappa=10^{-3}$ ($10^{-2}$) for the low (high) 
	Q cavity and $g_e\mu_B\Delta_B^\mathrm{max}=0.2\ns$. {\bf (b)} Fidelity $\mathcal{F}^{(1)}$ including 
	nuclear spin noise as a function of the degree of nuclear spin polarisation. The external field has been tuned to the optimal value 
	found numerically in (a). Line styles represent parameters as in (a). Circles and error bars indicate ensemble 
	averages and (25\%, 75\%) quantiles of the fidelity. {\bf(c)} Fidelity $\mathcal{F}^{(2)}$ with respect to the ideal 
	spin--two-photon state, obtained by scattering two photons on the QD with a time delay of $3t_0$.}
	\label{fig:fidelity}
\end{figure}

By first artificially setting the Overhauser field to zero, in Fig.~\ref{fig:fidelity}(a) we show how the spin--one photon Bell 
state fidelity $\mathcal{F}^{(1)}$ can be optimised by tuning the external magnetic field. We see that near-unity scattering fidelity 
is reached when the external field is approximately the cavity-enhanced QD linewdith, 
$b_\mathrm{ext}\simeq \Gamma_\mathrm{cav}=4g^2/\kappa$, as it is depicted in Fig.~\ref{fig:intro}(c). 
This ensures that an incoming photon has a high probability of scattering off one of the two 
possible transitions while also ensuring that they are adequately separated.
In this regime, the fidelity is limited by the finite bandwidth of the input photon, 
since any off-centre frequency components lead to an unevenly weighted superposition in the scattered state. 
In Fig.~\ref{fig:fidelity}(b) and (c), we show the fidelities $\mathcal{F}^{(1)}$ and $\mathcal{F}^{(2)}$  
including the nuclear environment, 
shown as a function of the nuclear environment polarisation, 
ranging from maximally unpolarised ($\Delta_B=\Delta_B^\mathrm{max}$)
to the fully polarised ($\Delta_B=0$) regime, 
and for high (red, dashed curve) and low (blue, solid) cavity $Q$-factors, 
corresponding to QDs with broad and narrow Purcell-enhanced transition lines. 
We see that even for an unpolarised nuclear environment, fidelities of the two-photon 
state are above $90\%$ for $Q=13000$. 
Higher $Q$-factors are advantageous since they correspond 
to larger QD linewidths and hence larger optimal external field strengths, which 
in turn mean the strength of the external field relative to the 
Overhauser field is greater. This results in increased stability of the QD eigenstructure and 
purity of the QD--photon scattering process, while also ensuring 
that the Overhauser field leads only to fluctuations in the magnitude of the field. 

We emphasise that the internal photon--QD interaction in the protocol 
is in principle deterministic, with the quantum efficiency being limited only by scattering of light into non-cavity modes, 
which is heavily suppressed in moderate to high $Q$-cavities~\cite{Iles-Smith2017Nature,somaschi2016near}. To obtain a 
purely polarisation-entangled state, however, it is necessary to erase the frequency degree of freedom in $\ket{\psi^{(n)}}$. 
This is an unavoidable consequence of the state's insensitivity to ensemble dephasing, and could be achieved, for example, 
using fast single-photon 
detectors~\cite{schaibley2012effect,economou2005unified} 
or ultra fast non-linear frequency converters~\cite{de2012quantum}.

\begin{figure}
	\centering
	\includegraphics[width=\columnwidth]{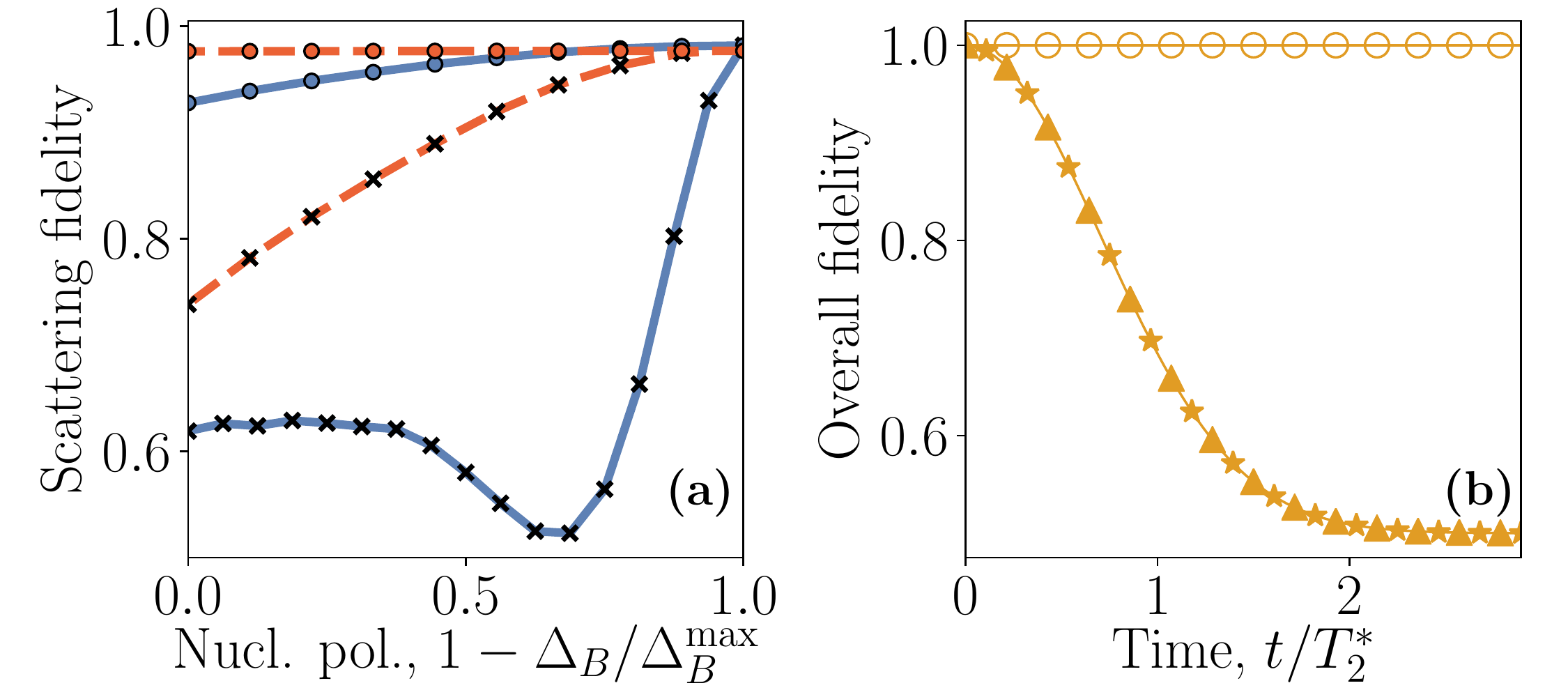}
	\caption{Comparison of the present scheme with existing Protocols A, B and C described in the main text. {\bf(a)} 
	Ensemble averaged spin--one-photon Bell state scattering fidelity $\mathcal{F}^{(1)}$ as in Fig.~\ref{fig:fidelity}(b) for protocol A (crosses) 
	and for the present scheme (circles, already shown in Fig. \ref{fig:fidelity}(b)).
	Red and blue lines correspond to high and low $Q$-factor parameters as in Fig.\ref{fig:fidelity}(b). 
	{\bf(b)} Ensemble averaged photonic state fidelity after spin 
	projection as a function of time, assuming unit scattering fidelity, shown for Protocols B (stars) and C (trianges). 
	Both 	show a rapid decay due to large spread of possible Overhauser fields, while the present scheme (circles) is unaffected 
	due to the form of Eq.~({\ref{eq:2qb-ket}}) which acquires only a global phase in time.}
	\label{fig:alternative-schemes}
\end{figure}

\section{Comparison to alternative protocols}
To benchmark our protocol, we compare it to three alternative existing schemes. 
The first scheme (Protocol A) is based on coherent scattering of single linearly polarised photons on a charged 
QD in the absence of an external field \cite{hu2008deterministic,PhysRevB.88.035315}. 
Using our noise model, we calculate the scattering fidelity of this protocol in the presence of the same realistic nuclear spin environment. 
The fidelity of generating the spin--one-photon Bell state $\mathcal{F}^{(1)}$ 
is shown in Fig.~\ref{fig:alternative-schemes}(a), where crosses indicate values using Protocol A, 
and the circles correspond to values using the present dephasing-resilient scheme (already shown in Fig. \ref{fig:fidelity}(b)). 
We see that the scattering fidelity of Protocol A is generally low, 
reaching values close to unity only for very high degrees of nuclear spin polarisation. 
The reason for this protocol's sensitivity to dephasing processes 
can be attributed to its lack of an 
external field, which leaves the QD eigenstructure highly exposed to Overhauser field fluctuations. 

The second alternative scheme we consider (Protocol B) is a multi-photon extension of the schemes used in 
Refs.~\cite{gao2012observation, schaibley2013demonstration, de2012quantum},  
that use emission of a QD in an external in-plane field. This protocol resembles the scheme we propose here,
but with the crucial difference that single photon scattering in our scheme is replaced by full $\pi$-pulse 
excitations followed by spontaneous emission. While the magnetic field does ensure stability of the 
spin eigenstructure and high scattering fidelity (unlike protocol A), the spectrally broad $\pi$-pulses mean 
energy is not conserved in all paths of the evolution. 
As we show in App.~\ref{sec:protocol-B}, the result is that the $n$-photon state contains terms which acquire phases that 
depend on the fluctuating total effective Zeeman energy $b^x=g_e\mu_B(B_\mathrm{ext}+B_\mathrm{N}^x)$ in different ways, and  
the state therefore loses its phase coherence on a short $T_2^*=\sqrt{2}/(g_e\mu_B\Delta_B)\sim \mathrm{ns}$ timescale, 
in much the same way as a single electron spin. 
Fidelities of a two-photon state obtained after excitation with two $\pi$-pulses followed by 
spin projection are shown in Fig.~\ref{fig:alternative-schemes}(b) with stars, 
where unit scattering fidelity is assumed, and $t$ represents time after spin projection. 
Also shown with triangles is the corresponding two-photon state fidelity for the linear cluster 
state generation proposal of Ref.~\cite{lindner2009proposal}, Protocol C, again assuming unit scattering fidelity. 
As in the case of Protocol B, this scheme is also sensitive to ensemble dephasing of the electron spin. 
The form of Eq.~({\ref{eq:2qb-ket}}), however, ensures the present scheme does not dephase by this mechanism, 
leading to coherence times well beyond nanoseconds, as shown by the open circles.

\comment{Is protocol B really like the machine gun?}
\comment{At the moment we don't say anything about Gershoni's paper, which we should.}

In summary, we have presented a spin-mediated multi-photon entanglement protocol which is robust against slow 
Overhauser field fluctuations, meaning that coherence is limited to the pure spin dephasing time 
$T_2$ of microseconds, rather than the inhomogeneous dephasing time $T_2^*$ 
of nanoseconds.
With a suitable frequency eraser, the protocol 
can be used as a source of high-fidelity three-photon GHZ states, 
which through linear optical operations can be transformed to a universal 
quantum resource for measurement-based quantum computing \cite{gimeno2015three}. 
We emphasise that no spin echo or nuclear polarisation techniques are necessary, 
and that optical excitation could be achieved with 
readily obtainable weak coherent laser pulses, or instead with 
narrowband single photons for deterministic operation.

\begin{acknowledgements}

We thank Anders S. S\o rensen, Thomas Nutz, Petros Androvitsaneas, Dale Scerri, and Erik Gauger for helpful discussions. E.V.D, J.I.-S., and J.M. 
acknowledge funding from the Danish Council for Independent Research (DFF-4181-00416). 
This project has received funding from the European Union'€™s Horizon 2020 research and innovation 
programme under the Marie Sk{\l}odowska-Curie grant agreement No.~703193.

\end{acknowledgements}

\appendix

\section{Description of model}
\label{sec:model}
We consider a singly negatively charged quantum dot (QD) in a one-sided cavity, which is driven by a
polarised weak optical pulse. The cavity field is resolved in two orthogonal circular polarisations with mode operators $a_+$ and $a_-$, satisfying $[a_\lambda,a^\dagger_{\lambda'}]=\delta_{\lambda\lambda'},\; [a_\lambda,a_{\lambda'}]=[a^\dagger_\lambda,a^\dagger_{\lambda'}]=0$. We assume that the cavity is resonant with the QD transition at a frequency of $\omega_0$. Further, the cavity is coupled to the optical electromagnetic environment, resolved in two polarisations with mode operators $b_{\lambda q}$, where $\lambda=\pm$ denotes the polarisation and $q$ denotes the mode index. 
As a basis for the QD, we use the spin eigenstates projected along the $z$-direction, taken as the optical axis. For the charged ground states, these are $\ket{\uparrow}$ and $\ket{\downarrow}$, while for the corresponding trion states they are $\ket{\Uparrow}$ and $\ket{\Downarrow}$, denoting the heavy-hole spin states with spin projection eigenvalues $J_z=\pm 3/2$. Due to isotropic strain, the light holes with $J_z=\pm 1/2$ are split off from the heavy holes by an energy, $\Delta_\mathrm{LH}$, much larger than the linewidth of the transition, and we can ignore them in the light-matter interaction \cite{fischer2008spin}. The QD is subject to a magnetic field, $\mathbf{B}$ in an arbitrary direction described by the polar (azimuthal) angle, $\theta$ ($\phi$), and with a magnitude of $B$.
Moving to a frame rotating with the resonance frequency, $\omega_0$, the total Hamiltonian can be written as $\hat{H}(t)=H_0(t) + H_B + H_\mathrm{EM}^0 + H_\mathrm{EM}^I$, with ($\hbar=1$)
\begin{align}
\begin{split}
H_0(t) &= \eta(t)\mathbf{e}_\mathrm{in}^\dagger\mathbf{A} + g\mathbf{\Sigma}^\dagger\mathbf{A} + \mathrm{H.c}, \\ H_B &= \mu_B\mathbf{B}\cdot(g_e\mathbf{S}_e - g_h\mathbf{S}_h), \\
H_\mathrm{EM}^0 &= \sum_{\lambda q} (\omega_q-\omega_0) b_{\lambda q}^\dagger b_{\lambda q}, \;\; H_\mathrm{EM}^I = \sum_{\lambda q} g_q b_{\lambda q} a_\lambda^\dagger + \mathrm{H.c.},
\end{split}
\end{align}
where $\mathbf{A}$ is the polarisation-resolved vectorial mode operator $(a_+, a_-)^\mathrm{T}$, $\mathbf{\Sigma}=(\dyad{\uparrow}{\Uparrow}, \dyad{\downarrow}{\Downarrow})^\mathrm{T}$ is the spin-resolved QD transition operator, $g$ is the QD--cavity coupling rate, $\mu_B$ is the Bohr magneton, $g_e$ ($g_h$) is the electron (hole) Land\'e factor, $\mathbf{S}_e$ ($\mathbf{S}_h$) is the vectorial spin operator for the electron (hole) subspace, $\omega_q$ is the frequency of the $q$'th environmental mode and $g_q$ is the coupling rate between the cavity and the $q$'th environmental mode. The pulse envelope, $\eta(t)$ is taken to be Gaussian,
$\eta(t)=\eta_0\exp[-(t/t_0)^2]$ and $\mathbf{e}_\mathrm{in}$ the input polarisation Jones vector in the circular basis. 

To write down a practical form of $H_\mathrm{B}$, we use the spin eigenstates as a basis. For the electron spin in the ground state manifold, we use the Zeeman eigenstates determined by the direction of the magnetic field, $\ket{\phi_+}=\cos\theta/2\ket{\uparrow} + e^{i\phi}\sin\theta/2\ket{\downarrow}, \; \ket{\phi_-}=e^{-i\phi}\cos\theta/2\ket{\uparrow} - \sin\theta/2\ket{\downarrow}$.
As for the trion spin, treating the magnetic field interaction perturbatively to first order in the parameter $\mu_B g_h B/\Delta_\mathrm{LH}$, the light and heavy hole manifolds remain uncoupled. The heavy hole eigenstates are then $\ket{\Downarrow}$ and $\ket{\Uparrow}$ with associated energies $\pm3/2\mu_B g_h B\cos\theta$. In this basis, $H_B$ takes the form
\begin{align}
\begin{split}
H_\mathrm{B} &= -\frac{3}{2}\tilde{g}_h b\cos\theta \qty(\dyad{\Uparrow} - \dyad{\Downarrow}) \\ &\hspace{0.1\columnwidth}+ \frac{b}{2}\qty(\dyad{\phi_+} - \dyad{\phi_-}),
\end{split}
\end{align}
with $b=\mu_Bg_eB$ and $\tilde{g}_h=g_h/g_e$.

The interaction with the electromagnetic environment can be simplified by applying a standard Born-Markov approximation, corresponding to assuming a flat spectral density over the relevant frequency range \cite{carmichael2009statistical1}. With this approximation and neglecting the environmentally induced Lamb shift, the perturbative master equation treating $H_\mathrm{EM}^I$ to second order is $\dot{\rho}(t)=\mathcal{L}(t)\rho(t)$ with $\mathcal{L}(t)$ the time-dependent Liouvillian, 
\begin{align}
\label{eq:master-eq}
\mathcal{L}(t) = -i[H_0(t) + H_\mathrm{B},\:\cdot\:] +
\kappa\sum_{\lambda=\pm} \qty(a_\lambda\cdot a_\lambda - \frac{1}{2}\{a_\lambda^\dagger a_\lambda,\cdot \}),
\end{align}
where $\kappa$ is the cavity dissipation rate. This master equation can be solved numerically to obtain the time evolution of the density operator \cite{johansson2012qutip}.

The polarisation resolved reflected output modes, $\xi_\lambda$, can be calculated from the cavity mode using input-output theory \cite{steck2007quantum} as $\xi_\lambda(t) = i\mathbf{e}_\lambda^\dagger\mathbf{e}_\mathrm{in}\frac{\eta(t)}{\kappa} + \mathbf{e}_\lambda^\dagger\mathbf{A}$ with $\mathbf{e}_\lambda$ the Jones polarisation vector describing the polarisation mode $\lambda$. The $H$ and $V$ polarisations are described by the Jones vectors $\mathbf{e}_H=\frac{1}{\sqrt{2}}(1,1)^\mathrm{T},\; \mathbf{e}_V=\frac{1}{\sqrt{2}}(1,-1)^\mathrm{T}$.

\section{Fidelity measures}
\label{sec:fidelity-measures}
With the full time evolution of the cavity-QD density operator, $\rho(t)$, at hand, we can calculate any properties of the system. In particular, we can calculate the fidelity of the composite state consisting of the polarisation of scattered light and the internal spin state of the QD. However, this fidelity cannot be evaluated directly from the time-resolved density operator. Care must be taken, because the light polarisation must be defined in terms of the reflected light from the cavity, described by the output field operators, $\xi_H(t)$ and $\xi_V(t)$.

First, we consider a single photon scattered on the QD. In general, the two-qubit space spanned by the QD spin and the polarisation of the scattered photon can be described by the basis $\mathcal{B}^{(1)}=\{\ket{H\phi_+},\ket{V\phi_-},\ket{H\phi_-},\ket{V\phi_+}\}$, with the superscript (1) signifying that the space spans the polarisation of one photon and the QD spin. We denote the true density matrix of the post-scattering state of the single photon--spin system in this basis by $\rho^{(1)}$. We denote by $\chi^{(1)}$ the ideal density operator corresponding to the pure state $\ket*{\psi^{(1)}_\mathrm{pure}}=\alpha\ket{H\phi_+}+\beta\ket{V\phi_-}$. In the basis $\mathcal{B}^{(1)}$ it takes the form
\begin{align}
\chi^{(1)} = \mqty(\abs*{\alpha}^2 & \alpha\beta^* & 0&0 \\ \alpha^* \beta & \abs*{\beta}^2 &0&0 \\ 0&0&0&0\\0&0&0&0).
\end{align}
The fidelity between two density operators, $\rho_1$ and $\rho_2$ takes the form $\mathcal{F}_{12}=\tr(\rho_1\rho_2)$, if at least one of the density operators is pure. In our case, $\chi^{(1)}$ is pure and we may write the fidelity as 
\begin{align}
\label{eq:bell-rho-fidelity}
\begin{split}
\mathcal{F}^{(1)} &= \abs*{\alpha}^2\rho^{(1)}_{11} + \abs*{\beta}^2\rho^{(1)}_{22} + 2\Re{\alpha\beta^*\rho^{(1)}_{21}},
\end{split}
\end{align}
which shows that we only need to calculate four matrix elements of $\rho^{(1)}$ to evaluate the fidelity. To evaluate these matrix elements, we define the joint spin-polarisation expectation values $\ev*{S^{P\lambda}_{P'\lambda'}}= \int\dd{t}\ev*{\xi_P^\dagger(t)\xi_{P'}(t)\sigma_{\lambda\lambda'}(t)}/\mathcal{N}^{(1)} $ with $M_{\lambda\lambda'}=\dyad{\phi_\lambda}{\phi_{\lambda'}}$ the input intensity normalisation $\mathcal{N}^{(1)}=\int_{-\infty}^\infty\dd{t} \xi_\mathrm{in}^*(t)\xi_\mathrm{in}(t)= \sqrt{\pi/2} \eta_0^2 t_0/\kappa^2$, obtained using  $\xi_\mathrm{in}(t)=i\eta(t)/\kappa$. This normalisation accounts for the fact that we model the incoming photon as a weak coherent pulse. We then find that the $\mathcal{F}^{(1)}$ can be calculated as
\begin{align*}
\mathcal{F}^{(1)}= \abs*{\alpha}^2 \ev*{S^{H+}_{H+}} + \abs*{\beta}^2\ev*{S^{H+}_{V-}} + 2\Re[\alpha^*\beta\ev*{S^{V-}_{V-}}].
\end{align*}

Now we shall consider the scattering of a second photon on the cavity after some time, $\tau$. In App. \ref{sec:n-photon}, we calculate the explicit state, but here we shall simply use the general form for the ideal scattered state
\begin{align}
\label{eq:cluster-out}
\begin{split}
\ket*{\psi^{(2)}_\mathrm{pure}} &= \alpha\ket{H_1 H_2 \phi_+} + \beta\ket{H_1 V_2 \phi_-} \\ &+ \gamma\ket{V_1 H_2 \phi_-} + \delta\ket{V_1 V_2 \phi_+},
\end{split}
\end{align}
with corresponding density operator $\chi^{(2)}=\dyad*{\psi^{(2)}}$.
Note that the coefficients $\alpha$ and $\beta$ are not those entering $\chi^{(1)}$.
In analogy with the single-photon scattering case, we shall denote the true post-scattering density operator by $\rho^{(2)}$. The fidelity becomes
\begin{align}
\begin{split}
\mathcal{F}&^{(2)} = \abs*{\alpha}^2\ev*{S^{HH+}_{HH+}} + \abs*{\beta}^2\ev*{S^{HV-}_{HV-}}  
+ \abs*{\gamma}^2\ev*{S^{VH-}_{VH-}} \\&+ \abs*{\delta}^2\ev*{S^{VV+}_{VV+}} + 2\Re\Big\{ 
\alpha^*\beta\ev*{S^{HV-}_{HH+}} + \gamma^*\delta \ev*{S^{VV+}_{VH-}} \\ &+ e^{ib^x\tau}\big[ \alpha^*\gamma \ev*{S^{VH-}_{HH+}} + \alpha^*\delta \ev*{S^{VV+}_{HH+}}
\\ &+ \beta^*\gamma \ev*{S^{VH-}_{HV-}} + \beta^*\delta\ev*{S^{VV+}_{HV-}} \big]
\Big\},
\end{split}
\end{align}
with 
\begin{align*}
\ev*{S^{PQ\lambda}_{P'Q'\lambda'}}=\frac{1}{{\mathcal{N}^{(2)}(\tau)}}&\int\dd{t} \langle \xi_P^\dagger(t)\xi_Q^\dagger(t+\tau)\\ & \sigma_{\lambda\lambda'}(t+\tau)\xi_{Q'}(t+\tau)\xi_{P'}(t)\rangle
\end{align*}
and 
\begin{align*}
\mathcal{N}^{(2)}(\tau) &= \int_{-\infty}^\infty \dd{t} \xi_\mathrm{in}^*(t)\xi_\mathrm{in}^*(t+\tau) \xi_\mathrm{in}(t+\tau)\xi_\mathrm{in}(t) 
\\
&=  \frac{1}{\kappa^4}\int_{-\infty}^\infty \dd{t} \abs{\eta(t)}^2\abs{\eta(t+\tau)}^2.
\end{align*}
Due to the symmetries of the proposed protocol as described in the main text, the fidelity turns out to be independent of the photon separation time, $\tau$.

\section{Multi-photon entanglement structure}
\label{sec:n-photon}
\subsection{Unitary dynamics}
\label{sec:unitary-dynamics}
The interaction between a string of $n$ photons and the QD can be entirely described by the unitary scattering operator, $U$, describing the asymptotic composite state resulting from of a single-photon scattering event. To find $U$, we have numerically calculated the post-scattering state of four orthogonal initial conditions using the methods described in Appendices \ref{sec:model} and \ref{sec:fidelity-measures},
\begin{align}
\label{eq:input-output-scattering}
\begin{split}
\ket{H,\omega_0}\ket{\phi_+} &\xrightarrow{U} \frac{1}{\sqrt{2}}\qty(\ket{H,\omega_0}\ket{\phi_+} - i\ket{V,\omega_0+b^H}\ket{\phi_-}), \\
\ket{V,\omega_0}\ket{\phi_+} &\xrightarrow{U} \frac{1}{\sqrt{2}}\qty(\ket{V,\omega_0}\ket{\phi_+} - i\ket{H,\omega_0+b^H}\ket{\phi_-}), \\
\ket{H,\omega_0}\ket{\phi_-} &\xrightarrow{U} \frac{1}{\sqrt{2}}\qty(\ket{H,\omega_0}\ket{\phi_-} + i\ket{V,\omega_0-b^H}\ket{\phi_+}), \\
\ket{V,\omega_0}\ket{\phi_-} &\xrightarrow{U} \frac{1}{\sqrt{2}}\qty(\ket{V,\omega_0}\ket{\phi_-} + i\ket{H,\omega_0-b^H}\ket{\phi_+}).
\end{split}
\end{align}
To establish the full unitary operator, we would need to find the evolution of initial conditions with photon frequencies $\omega_0 \pm b$ as well. However, to this end we are only interested in the scattering dynamics of photons resonant with the zero-field QD transition at $\omega_0$. In particular, when restricting the discussion to $H$-polarised input photons, we only need to know how $U$ works on $\ket{H,\omega_0}\ket{\phi_\pm}$. We then write the total scattered state as 
\begin{align}
\ket*{\psi^{(n)}} = \qty(\prod_j U_j)\ket{H,\omega_0}_1\cdots\ket{H,\omega_0}_n\ket{\phi_+},
\end{align}
where $U_j$ acts on the $j$'th photon and the QD.
In particular, for two photons, we obtain the state in Eq. (1) of the main text.
Generally speaking, the $n$-photon entangled state has the form $\ket*{\psi^{(n)}}=\frac{1}{\sqrt{2}}(\ket{n,+}\ket{\phi_+} + \ket{n,-}\ket{\phi_-})$. Here, $\ket{n,+}$ contains all superpositions of polarisation permutations with an even number of $y$-polarised photons, where each term in the superposition has a total photonic energy of $n\omega_0$. Similarly, $\ket{n,-}$ contains all terms with an odd number of $y$-polarised photons and all terms have a photonic energy of $n\omega_0+b^x$.

\subsection{Entanglement structure of spin--multi-photon state}
If we assume that the photon frequency degree of freedom is erased and can be factored out of the remaining state, the generating scattering transformation, \eqref{eq:input-output-scattering}, becomes non-unitary and takes the form $G_j=\frac{1}{\sqrt{2}}(\mathbb{1}_\mathrm{QD}\otimes\mathbb{1}_j - Y_\mathrm{QD}\otimes X_j)$, with $Y_\mathrm{QD}=i(\dyad{\phi_-}{\phi_+}-\dyad{\phi_+}{\phi_-})$ and $X_j=\dyad{H}{V}_j+\dyad{V}{H}_j$. Using this form of the scattering operator, we can write down the $n$-photon--spin entangled state. To do so, we change notation by defining the computational basis for the photon polarisation as $\ket{H}_k=\ket{0}_k,\; \ket{V}_k=\ket{1}_k$ and  $\ket{\phi_+}=\ket{0},\; \ket{\phi_-}=\ket{1}$ for the QD spin. The ket subscripts $k=1,\cdots,n$ shall be used for the photonic qubits, while $k=0$ denotes the spin qubit. We shall use $i_k\in \{0,1\} $ to denote the value of a qubit in the computational basis, $\mathbf{i}=(i_1, \cdots, i_m)$ denotes an $m$-bitstring and $\ket{\mathbf{i}}_\mathcal{S}$ the corresponding state with respect to the qubits in the ordered set $\mathcal{S}$. Further, we shall neglect normalisation factors for ease of notation. The $n$-photon scattered state can then be written as
\begin{align}
\begin{split}
\ket*{\psi^{(n)}}&=\prod_j G_j \ket{0,\cdots,0}_{\{0,\cdots, n\}} \\ &= \ket{0}_0\sum_{\mathbf{i}\in S_e(n)} \ket{\mathbf{i}}_{\{1,\cdots, n\}}
-i \ket{1}_0\sum_{\mathbf{i}\in S_o(n)} \ket{\mathbf{i}}_{\{1,\cdots, n\}},
\end{split}
\end{align}
with $S_e(n)=\{\ket{i_1,\cdots,i_n}|\sum_k i_k=2m, m\in \mathbb{N} \}$ and  $S_o(n)=\{\ket{i_1,\cdots,i_n}|\sum_k i_k=2m+1, m\in \mathbb{N} \}$. By singling out the $k$'th, $l$'th and $m$'th photonic qubits from the sums, we can rewrite this state as 
\begin{align}
\label{eq:c-decomposition}
\ket*{\psi^{(n)}} = \ket{c'_+}_{klm}\ket{R_+} + \ket{c'_-}_{klm}\ket{R_-},
\end{align} with the $k,l,m$-qubit states $\ket{c'_+}=\ket{000}+\ket{110}+\ket{011}+\ket{101}$, $\ket{c'_-}=\ket{001}+\ket{111}+\ket{010}+\ket{100}$ and the residual qubit states
\begin{align}
\begin{split}
\ket{R_+}&=\ket{0}_0\sum_{\mathbf{i}\in S_e(n-3)}\ket{\mathbf{i}}_{\{1,\cdots,n \}\backslash\{k,l,m\}} \\ &- i\ket{1}_0 \sum_{\mathbf{i}\in S_o(n-3)}\ket{\mathbf{i}}_{\{1,\cdots,n \}\backslash\{k,l,m\}}, \\
\ket{R_-}&=\ket{0}_0\sum_{\mathbf{i}\in S_o(n-3)}\ket{\mathbf{i}}_{\{1,\cdots,n \}\backslash\{k,l,m\}} \\ &- i\ket{1}_0 \sum_{\mathbf{i}\in S_e(n-3)}\ket{\mathbf{i}}_{\{1,\cdots,n \}\backslash\{k,l,m\}}.
\end{split}
\end{align}
The states $\ket{c'_\pm}_{klm}$ are local unitary equivalent to three-qubit linear cluster states, which are local unitary equivalent to three-photon GHZ states \cite{greenberger1989going}. This is seen by applying the Hadamard transformation, $\mathcal{H}=\dyad{0}+\dyad{1}{0}+\dyad{0}{1}-\dyad{1}$ to the $l$'th photon, $\mathcal{H}_l\ket{c'_\pm}_{\{klm\}}=\ket{c_\pm}_{\{klm\}}$ where $\ket{c_\pm}$ are the two orthogonal cluster states $\ket{000}\pm \ket{111}+\ket{100}\mp \ket{110}+\ket{001}\mp \ket{011}+\ket{101}\pm \ket{111}$.
From this form, we can easily calculate all single-qubit reduced density operators, which all take the form $\rho_k=\mathbb{1}$. Furthermore, we can calculate the two-qubit reduced density operators, which for two photonic qubits, $kl$, take the form $\rho_{kl}=\dyad{\mathcal{B}_1} + \dyad{\mathcal{B}_2}$ with the two Bell states $\ket{\mathcal{B}_1}=\ket{00}+\ket{11},\; \ket{\mathcal{B}_2}=\ket{01}+\ket{10}$. Two-qubit reduced density operators involving the spin qubit, take the similar form $\rho_{0k}=\dyad{\mathcal{B}'_1}+\dyad{\mathcal{B}'_2}$ with the rotated Bell states $\ket{\mathcal{B}'_1}=\ket{00}-i\ket{11},\; \ket{\mathcal{B}'_2}=\ket{01}-i\ket{10}$. From these reduced density operators, we can show that the generated state is not local unitary equivalent to a linear cluster state for more than three qubits. This is due to the necessary condition for local unitary equivalence that all reduced density operators must also be local unitary equivalent \cite{kraus2010local}. Since for a linear cluster state with more than three qubits there exist indices $kl$ such that $\rho_{kl}=\mathbb{1}$, the two states cannot be local unitary equivalent. 

However, from \eqref{eq:c-decomposition}, we can infer that performing local projective measurements on \emph{any} $n-3$ photons and the spin in the computational basis leaves the remaining three photons in an entangled state that is local unitary equivalent to a three-qubit GHZ or linear cluster state. This also holds in the particular case, where there are only three photons in the scattered state, and a projective measumerent is performed on the spin. A similar series of local measurements on $n-2$ photons and the spin leaves the remaining two photons in a Bell state, which is maximally entangled. Since these properties do not depend on the indices of the photonic qubits, we infer that the localisable entanglement is maximal and the entanglement length is infinite \cite{popp2005localizable}.

\section{Analysis of Protocol B}
\label{sec:protocol-B}
A protocol very similar to the one proposed in the main text has been used for generation of entanglement between a single photon and a QD  \cite{gao2012observation, schaibley2013demonstration, de2012quantum}. Here, the QD is initialised in the $\ket{\phi_+}$ ground state and excited to $\frac{1}{\sqrt{2}}(\ket{\Uparrow}+\ket{\Downarrow})$ by an $H$-polarised $\pi$-pulse resonant with the transition $\ket{\phi_+}\leftrightarrow\frac{1}{\sqrt{2}}(\ket{\Uparrow}+\ket{\Downarrow})$ at $\omega_0-b^x/2$. As this state decays, a photon is emitted, which is entangled with the spin of the QD,
\begin{align}
\ket*{\psi^{(1)}}=c_H\ket{H,\omega_0-b^x/2}_1\ket{\phi_+}+c_V\ket{V,\omega_0+b^x/2}_1\ket{\phi_-}
\end{align}
with $\abs{c_i}^2=1/2$.
This state is protected against dephasing, because both terms in the superposition have the same total energy of $\omega_0$. To add a second photon to the state, the QD is excited again. This time, we have to use a two-colour $\pi$-pulse, because the QD is in a superposition of the two ground states. Immediately after the excitation, the system is in the state
\begin{align}
c_H\ket{H,-b^x/2}_1\frac{\ket{\Uparrow}+\ket{\Downarrow}}{\sqrt{2}}+c_V\ket{V,+b^x/2}_1\frac{\ket{\Uparrow}-\ket{\Downarrow}}{\sqrt{2}},
\end{align}
where we have transformed to a frame rotating with $\omega_0$.
As the QD decays, the state becomes
\begin{align}
\begin{split}
\ket*{\psi^{(2)}} =\: &c_{HH}\ket{H,-b^x/2}_1\ket{H,-b^x/2}_2\ket{\phi_+} \\+ &c_{HV}\ket{H,-b^x/2}_1\ket{V,+b^x/2}_2\ket{\phi_-}  \\ 
+&c_{VH}\ket{V,+b^x/2}_1\ket{H,+b^x/2}_2\ket{\phi_-} \\+ 
&c_{VV}\ket{V,+b^x/2}_1\ket{V,-b^x/2}_2\ket{\phi_+},
\end{split}
\end{align}
with $\abs{c_{\alpha\beta}}^2=1/4$
Here, the two first terms have an energy of $-b^x/2$, whereas the two last terms have an energy of $+b^x/2$. In the time until the next excitation event, $\tau$, the state will evolve freely. Recalling that $b^x=b_\mathrm{ext}^x+b_\mathrm{N}^x$, the time evolution is
\begin{align}
\begin{split}
\ket*{\psi^{(2)},\tau}& =  \\ &e^{+i(b_\mathrm{ext}^x+b_\mathrm{N}^x) \tau/2} \Big[c_{HH}\ket{H,-b^x/2}_1\ket{H,-b^x/2}_2\ket{\phi_+} \\ &\hspace{0.2\columnwidth}+ c_{HV}\ket{H,-b^x/2}_1\ket{V,+b^x/2}_2\ket{\phi_-}\Big]  \\ 
+ &e^{-i(b_\mathrm{ext}^x+b_\mathrm{N}^x) \tau/2}\Big[ c_{VH}\ket{V,+b^x/2}_1\ket{H,+b^x/2}_2\ket{\phi_-} \\ &\hspace{0.2\columnwidth}+ 
c_{VV}\ket{V,+b^x/2}_1\ket{V,-b^x/2}_2\ket{\phi_+}\Big].
\end{split}
\end{align}
The fidelity with respect to $\ket*{\psi^{(2)}}$ is $\abs{\ip*{\psi^{(2)}}{\psi^{(2)},\tau}}^2 = \frac{1}{2}\qty{1+\cos[(b^x_\mathrm{ext}+b^x_\mathrm{N})\tau]}$. On performing an ensemble average over the weight distribution of the Overhauser field, the fidelity becomes $\mathcal{F}=\frac{1}{2}\int_{-\infty}^\infty\dd{b^x_\mathrm{N}}w(b^x_\mathrm{N};\delta_b)\qty{1+\cos[(b^x_\mathrm{ext}+b^x_\mathrm{N})\tau]} = \frac{1}{2}\qty{1+e^{-(\tau/T_2^*)^2}\cos(b_\mathrm{ext}^x \tau)}$, with $T_2^*=\sqrt{2}/(g_e\mu_B\Delta_B)$. Such dephasing processes will take place between all of the following excitation events. The time between excitations is limited by the lifetime of the QD, and if we assume that this is much shorter than the coherence time, $T_2^*$, we may neglect dephasing between excitations for a few photons. However, after spin projection, the photonic state will be subject to dephasing of the same nature. Measuring the spin in the basis $\{\phi_+,\phi_-\}$ leaves the two emitted photons in either of the two states
\begin{align}\begin{split}
\ket*{\psi_+^{(2)}} &= \sqrt{2}\ip*{\phi_+}{\psi^{(2)}} \\ &= \sqrt{2}\Big[ c_{HH}\ket{H,-b^x/2}_1\ket{H,-b^x/2}_2 \\ & \hspace{0.1\columnwidth}+ c_{VV}\ket{V,+b^x/2}_1\ket{V,-b^x/2}_2\Big], \\
\ket*{\psi_-^{(2)}} &= \sqrt{2}\ip*{\phi_-}{\psi^{(2)}} \\ &= \sqrt{2}\Big[c_{HV}\ket{H,-b^x/2}_1\ket{V,+b^x/2}_2 \\ &\hspace{0.1\columnwidth} + c_{VH}\ket{V,+b^x/2}_1\ket{H,+b^x/2}_2\Big].
\end{split}\end{align}
After the projective measurement, the states evolve as 
\begin{align}\begin{split}
\ket*{\psi_+^{(2)},t} &= \sqrt{2}\ip*{\phi_+}{\psi^{(2)}} \\ 
&= \sqrt{2}\Big[c_{HH}e^{+i(b_\mathrm{ext}^x+b_\mathrm{N}^x)t}\ket{H,-b^x/2}_1\ket{H,-b^x/2}_2 \\ &\hspace{0.1\columnwidth}+ c_{VV}\ket{V,+b^x/2}_1\ket{V,-b^x/2}_2\Big], \\
\ket*{\psi_-^{(2)},t} &= \sqrt{2}\ip*{\phi_-}{\psi^{(2)}} = \sqrt{2}\Big[c_{HV}\ket{H,-b^x/2}_1\ket{V,+b^x/2}_2 \\ &+ c_{VH}e^{-i(b_\mathrm{ext}^x+b_\mathrm{N}^x)t}\ket{V,+b^x/2}_1\ket{H,+b^x/2}_2\Big].
\end{split}\end{align}
The fidelity of these states with respect to the $\ket*{\psi^{(2)}_\pm}$ is $f_\pm =\abs*{\ip*{\psi_\pm^{(2)}}{\psi_\pm^{(2)},t}}^2$. Since the outcome of the projective spin measurement is $\ket{\phi_+}$ and $\ket{\phi_-}$ with equal probability, the average fidelity is $\frac{1}{2}(f_++f_-)$. When performing an ensemble average over the Overhauser weight distribution, the resulting fidelity is $\mathcal{F}^{(2)}=\frac{1}{2}\int_{-\infty}^\infty\dd{b_\mathrm{N}^x} w(b_\mathrm{N}^x;\delta_b)(f_++f_-)=\frac{1}{2}\qty{1+e^{-(t/T_2^*)^2}\cos(b_\mathrm{ext}^x t)}$. In conclusion, the fidelity of the photonic state after spin projection decays with a time scale of $T_2^*$. This calculation can straightforwardly be extended to cover the spin-projected three-photon state, yielding a fidelity of $\mathcal{F}^{(3)}=\frac{1}{8}\qty{3+4e^{-(t/T_2^*)^2}\cos(b_\mathrm{ext}^x)+e^{-(2t/T_2^*)^2}\cos(2b_\mathrm{ext}^x)}$.

\end{document}